\documentclass[aps,showpacs,superscriptaddress]{revtex4}
\usepackage{color}
\usepackage{amssymb}
\usepackage{epsfig}
\usepackage{graphicx}
\usepackage{pstricks}
\usepackage{pst-coil}
\DeclareMathAlphabet{\mathsc}{OT1}{cmr}{m}{sc}

\newcommand{\Addrroma}{%

   Dipartimento di Fisica "E. Amaldi", Universit\'a degli Studi Roma Tre \\

   Via della Vasca Navale 84, 00146 Rome, Italy}
\newcommand{\Addrcern}{%

CERN, Theory Division, 1211 Geneva 23, Switzerland}

\relax

\begin{document}

\begin{flushright}
RM3-TH/12-6
\end{flushright}
\title{Fine-tuning and naturalness issues in the two-zero neutrino mass textures}

\author{G.~Blankenburg} \email{blankenburg@fis.uniroma3.it}

\affiliation{\Addrroma}\affiliation{\Addrcern}

\author{D.~Meloni} \email{meloni@fis.uniroma3.it} 

\affiliation{\Addrroma}

\date{\today}

\begin{abstract}

In this paper we analyze the compatibility of two-zero neutrino Majorana textures with the recent experimental data. 
Differently from previous works, we use the experimental data to fix the values of the non-vanishing mass matrix entries and study 
in detail the correlations and degree of fine-tuning among them, which is also a measure of how naturally a given texture is able to describe
all neutrino data. This information is then used to expand the 
textures in powers of the Cabibbo angle; extracting random ${\cal O}(1)$ coefficients, we show that only in few cases 
such textures reproduce the mixing parameters in their 3$\sigma$ ranges.

\end{abstract}

\pacs{
11.30.Hv       % Flavor symmetries
14.60.-z       % Leptons
14.60.Pq       % Neutrino mass and mixing
}

\maketitle

\section{Introduction}
The recent T2K \cite{T2K}, MINOS \cite{minos}, Daya Bay  \cite{db}  and RENO \cite{collaboration:2012nd} results contain a strong indication 
in favor of a large reactor angle $\theta_{13}$, as derived from the analysis of 
$\nu_\mu \to \nu_e$ appearance and  $\bar \nu_e \to \bar \nu_e$ disappearance
flavour transitions. In particular, Daya Bay  and RENO  provide a clear evidence of several standard deviations from zero of the reactor angle:
\begin{eqnarray}
\label{res}
 \sin^2 2\theta_{13} = 0.092 \pm 0.016 \pm 0.005\;\text{(Daya Bay)} \qquad \sin^2 2\theta_{13} = 0.103 \pm 0.013  \pm 0.011 \;\text {(RENO)}\,.
\end{eqnarray}
This has recently pushed the model building industry to study the possibility to accommodate a large  $\theta_{13}$, together with the well
established solar and atmospheric mixing angles \cite{Meloni:2012ci}.
In spite of the good accuracy in the mixing parameter determination, the structure of the neutrino mass matrix is still 
unknown; two-zero textures of a Majorana type have been largely studied before the T2K and MINOS data 
\cite{others}  and an update analysis (including the hints for a large $\theta_{13}$) has been recently presented
in \cite{Fritzsch:2011qv} and \cite{Ludl:2011vv}.
The common approach of such studies is that of using the zeros of the Majorana  mass matrix to get simple relations among
the neutrino mixing angles and mass differences, independently on the parameters defining the mass matrices. 
Such relations can then be used to determine the three neutrino masses and three CP-violating phases. It turns out that 
some of the two-zero textures are still compatible with the recent data at the 3$\sigma$ level whereas
all the textures marginally compatible or incompatible with the pre-T2K and MINOS results are now definitely ruled out 
\cite{Fritzsch:2011qv}. In addition, a look at Fig.11 of the same Ref.\cite{Fritzsch:2011qv} shows that also the texture called
$B_4$ (see later for its explicit form) has a problem with $\theta_{13}$, since its distribution at the 3$\sigma$ level does not
touch the value $\theta_{13}=7^\circ$ and is then marginally compatible with eq.(\ref{res}).

Although the procedure outlined above is very powerful to understand how the mixing angles and phases are related to each others 
(for a given texture), it gives no hint on the order of magnitude of the parameters defining the mass matrices; it also does not 
reveal whether the matrix elements are correlated and/or if they have to be fine-tuned (and to what degree) to reproduce the 
experimental data.  In this letter we want to clarify all these issues and study in details the features of the mass matrix elements
needed to accommodate the experimental results within a given texture. In principle one can use different approaches; here 
we rely on two different {\it estimators}. We first perform a $\chi^2$ analysis of the still allowed two-zero textures, with the aim
of getting a global information on the goodness of the mass matrix to fit the neutrino data. 
For every texture, we explicitly show  the values of the parameters that fit the data, in
such a way that the reader can easily notice the deviation of the mass matrix parameters from the natural 
(in absence of a more fundamental flavour model) ${\cal O}(1)$ value; such a deviation gives an idea of the naturalness 
of the textures.

Beside the minimum of the $\chi^2$, 
the fit procedure will also return the corresponding values of 5 neutrino observables, namely  the two independent mass differences 
$\Delta m^2_{sol}$ and $\Delta m^2_{atm}$ and the three mixing angles $\theta_{13}$, $\theta_{12}$ and $\theta_{23}$. 
We do not include the Dirac CP phase into the $\chi^2$ definition since its value has not been measured yet.
The minimum of the $\chi^2$ cannot be the only criterion to classify mass models, since a relatively good value can be obtained at the 
prize of a strong fine-tuning among the mass parameters. For this reason, we also 
give for each minimum a quantitative measurement (our second estimator) of the amount of fine-tuning needed. 
This is obtained calculating, for every parameter, the shift from the best fit value that changes the $\chi^2$ by one unit, 
with all other parameters fixed at their best fit values. 

Since our analysis is intended to make a relative comparison for the degree of fine-tuning between
various textures, we do not take into account the larger fine-tuning implied by the tiny errors on the masses in the charged lepton sector, which would 
otherwise make our comparison completely useless.

The results of such a procedure for a normal order (NH) of the neutrino mass eigenstates is described in Section \ref{text}. 
Using the best fit values of the mass matrix parameters, we expand the textures
in terms of the Cabibbo angle $\lambda \equiv \lambda_C$, with generic complex coefficient with absolute values of ${\cal O}(1)$ for any non-vanishing matrix elements. 
In this way, we automatically take into account the hierarchy among different entries of the mass matrix but it is not obvious that 
these coefficients combine to reproduce the best fit points of the neutrino observables. It is then interesting to 
study separately the degree of predictivity of such textures, extracting randomly the complex coefficients and analyzing the values of
the obtained mass differences and mixing angles. This procedure and its outcomes are detailed in Section \ref{textpre}, while in Section \ref{inverted}
we repeat the same exercise for the inverted hierarchy.
Section \ref{concl} is devoted to our conclusions.

\section{Observables and analysis details}
\label{text}
Although there exist a large class of Majorana mass matrices with two-zero entries, we restrict our analysis to the seven models still compatible with the data, according 
to \cite{Fritzsch:2011qv}, from which we also  adopt here the same nomenclature:

\begin{eqnarray}
\label{aini}
A^{}_1: ~~ \left(\matrix{ 0 & 0 & a_1 \cr 0 & a_2 &
a_3 \cr a_1 & a_3 & a_4 \cr} \right) \, , ~~~~ 
A^{}_2: ~~ \left(\matrix{ 0 & a_1 & 0 \cr a_1 & a_2 &
a_3 \cr 0 & a_3 & a_4 \cr} \right) \, ;
\end{eqnarray}
%and
\begin{eqnarray}
B^{}_1: ~~ \left(\matrix{ a_1 & a_2 & 0 \cr a_2 & 0 &
a_3 \cr 0 & a_3 & a_4 \cr} \right) \, , \qquad 
B^{}_2: ~~
\left(\matrix{ a_1 & 0 & a_2 \cr 0 & a_3 & a_4 \cr
a_2 & a_4 & 0 \cr} \right) \, , ~~~~~
\qquad
B^{}_3: ~~ \left(\matrix{ a_1 & 0 & a_2 \cr 0 & 0 &
a_3 \cr a_2 & a_3 & a_4 \cr} \right) \, , \qquad 
B^{}_4: ~~ \left(\matrix{ a_1 & a_2 & 0 \cr a_2 & a_4 &
a_3 \cr 0 & a_3 & 0 \cr} \right) \, 
\end{eqnarray}
%and
\begin{eqnarray}
\label{ci}
C: ~~ \left(\matrix{ a_1 & a_2 & a_3 \cr a_2 & 0 &
a_4 \cr a_3 & a_4 & 0 \cr} \right) \; .
\end{eqnarray}
A general complex symmetric matrix contains 6 moduli and 6 phases, for a total of 12 independent parameters. 
The redefinition of the neutrino fields allows to eliminate three phases; other two can be thrown away if they are
associated with the vanishing elements of the mass matrix. In total, for two-zero textures we are left with 4 moduli and
one phase $\varphi$ that, as it can be easily demonstrated with a trivial computation, can always be associated to $a_4$, 
for a total of 5 independent parameters. 
These are the quantities to be determined by the $\chi^2$ fit.

It has been pointed out in \cite{Fritzsch:2011qv} that the zeros in $A_{1}-A_{2}$ as well as those in 
$B_{1}-B_{2}$ and $B_{3}-B_{4}$ are related by a $P_{23}$ symmetry
\begin{equation}
P_{23}= \left(\matrix{ 1 & 0 & 0 \cr 0 & 0 &
1 \cr 0 & 1 & 0 \cr} \right)
\end{equation}
in such a way that:
\begin{equation}
 A_2 = P_{23}\,A_1\,P_{23}^T
\end{equation}
and so on. This in turn implies:
\begin{equation}
\label{p23}
\theta_{12}^{A_2}=\theta_{12}^{A_1}\qquad\theta_{13}^{ A_2}=\theta_{13}^{A_1}
\qquad\theta_{23}^{A_2}=\frac{\pi}{2}-\theta_{23}^{ A_1}\qquad \delta_{CP}^{ A_2}=\delta_{CP}^{ A_1}\,,
\end{equation}
and similarly for the pairs $B_{1,2}$ and $B_{3,4}$.
Analytical as well as numerical considerations on the range of the mixing parameters implied by the previous textures 
have been already analyzed in the literature (see for example \cite{others}); the main conclusions are that the textures $A_i$ 
describe the data much better in normal hierarchy than in the inverted one whereas  the texture C allows for the inverted hierarchy.
For the $B_i$'s, the mass spectrum is quasi-degenerate and it depends on the assumed octant of $\theta_{23}$; for $\theta_{23} < \pi/4$
$B_{1,3}$ are compatible with the normal hierarchy and $B_{2,4}$ with the inverted one (the behavior is opposite for $\theta_{23}$ in the second octant).

In order to quantify the qualitative behaviors described above, we start with
the normal hierarchy (NH) case, using the best-fit points and the 1$\sigma$ uncertainties on the mixing parameters from \cite{Fogli},
summarized in Tab.\ref{data}. 
% Our choice is mainly motivated by the fact that the fitted $\theta_{13}$ is large and compatible with eq.(\ref{res}), so 
% a possible updated analysis including the Daya Bay and RENO results would not change our conclusions dramatically.
\begin{table}[b!]
\begin{center}
\begin{tabular}{cccccc}
  \hline
  \hline
  Parameter & $\Delta m^2_{sol}~(10^{-5}~{\rm eV}^2)$ & $\Delta m^2_{atm}~(10^{-3}~{\rm eV}^2)$
  & $\sin^2 \theta_{12}$ & $\sin^2\theta_{23}$ & $\sin^2\theta_{13}$ \\
  \hline
  Best fit (NH) & $7.50$ & $2.47$ & $0.30$ & $0.41\,\oplus\,0.59$ & $0.023$ \\
  $1\sigma$ & $[7.32, 7.69]$ & $[2.40, 2.54]$ & $[0.287, 0.313]$
  & $[0.385, 0.447]\,\oplus\,[0.568,0.611]$ & $[0.0207, 0.0253]$ \\
  \hline
   Best fit (IH) & $7.50$ & $-2.43$ & $0.30$ & $0.41\,\oplus\,0.59$ & $0.023$ \\
  $1\sigma$ & $[7.32, 7.69]$ & $[-2.39, -2.50]$ & $[0.287, 0.313]$
  & $[0.385, 0.447]\,\oplus\,[0.568,0.611]$ & $[0.0207, 0.0253]$ \\
  \hline 
\end{tabular}
\end{center}
\caption{\label{data}\it The latest global-fit results from \cite{Fogli}.}
\end{table}
The definition of the $\chi^2$ is as follows:
\begin{eqnarray}
\chi^2 = \Sigma_{{ij}=12,13,23} \frac{\left[\sin^2 \theta_{ij}-(\sin^{2} \theta_{ij})^{data}\right]^2}{\sigma^2_{ij}} + 
\Sigma_{k=sol,atm} \frac{\left[\Delta m^2_{k}-(\Delta m^{2}_{k})^{data}\right]^2}{\sigma^2_k}\,,
\end{eqnarray}
where $(\sin^2 \theta_{ij})^{data}$ and $(\Delta m^{2}_{k})^{data}$ are the values of the mixing angles and mass differences reported in Tab.\ref{data}, respectively,
and $\sigma$ the corresponding 1$\sigma$ errors also quoted in Tab.\ref{data};
 $\sin^2 \theta_{ij}$ and $\Delta m^{2}_{k}$ depend on the five unknowns $a_{1,..,4}$ and $\varphi$.
Given that two statistically equivalent best fit values for $\theta_{23}$ emerged from the analysis in \cite{Fogli}, 
for every textures we fit both values and retained the  $a_{1,..,4}$ and $\varphi$ corresponding to the 
minimum among the two $\chi^2$.

To estimate the degree of fine-tuning we used the parameter $d_{FT}$ defined in \cite{Blankenburg}. 
This adimensional quantity is obtained as the sum of the absolute values of the ratios between each parameter and its "error", 
defined for this purpose as the shift from the best fit value that changes the $\chi^2$ by one unit, 
with all other parameters fixed at their best fit values (this is not the error given by the fitting procedure because in that case 
all the parameters are varied at the same time and the correlations are taken into account):
\begin{equation}
\label{fine-tuning}
d_{FT} = \sum \left |\frac{par_i}{err_i} \right|\,.
\end{equation}
It is clear that $d_{FT}$ gives a rough idea of the amount of fine-tuning involved in the fit because if some $|err_i/par_i|$ are very small it means that it 
takes a minimal variation of the corresponding parameter to make a large difference on the $\chi^2$. As in \cite{Blankenburg}, we can compare this quantity 
with the sum of the absolute values of the ratios between each observable and its error as derived from the data:
\begin{equation}
\label{data-precision}
d_{Data} = \sum \left |  \frac{obs_i}{err_i}\right|\,,
\end{equation}
that for the set of data in Tab.\ref{data} is $d_{Data}\sim 100$.

\vskip .3 cm
We have summarized our numerical results in Tab.\ref{fit}, where we have shown the value of the $\chi^2_{min}$, the fine-tuning parameter $d_{FT}$, the best fit 
values of the neutrino matrix elements and the predictions of some of the corresponding experimental observables. 
Notice that we did not display the values at the best fit $(a_i,\varphi)$ for $\theta_{12}$ and 
the mass differences, since we always got values indistinguishable from Tab.\ref{data}. 
\begin{table}[h!]
\begin{center}
\begin{tabular}{cccccccccc}
  \hline
  \hline
  texture & $\chi^2_{min}$ & $d_{FT}$ & $\sin^2 \theta_{13}$  &  $\sin^2\theta_{23}$ & $a_1$ & $a_2$ & $a_3$ & $a_4$ & $\varphi$\\
  \hline
$A_1$ & $8.5\cdot 10^{-11}$& $2.2\cdot 10^{2}$  & .023  &  .41   & 0.99 & 2.2 & -2.3& 2.8 &- 0.43  \\
$A_2$ &  $1.3 \cdot 10^{-11}$ &$2.2\cdot 10^{2}$ &  .023 &   .59 & -0.99 & 2.8 & -2.3  & 2.2& -0.43\\
$B_1$ & $5.8\cdot 10^{-9}$&$5.3\cdot 10^{3}$ &  .023  &  .41 & -4.5& -0.52 & 5.4 &  2.1& -3.3\\
$B_2$ & $8.3\cdot 10^{-7} $ &$5.2\cdot 10^{3}$ & .023 & .59 & 4.5 & -0.52 & -2.1 &5.5 & 1.5\\
$B_3$ & $3.9\cdot 10^{-10}$ & $6.1\cdot 10^{3}$ & .023 & .41 & -4.8&  0.41 & 5.7 & -2.0& 0.27 \\
$B_4$ & $1.4\cdot 10^{-8}$ &$5.7\cdot 10^{3}$ & .023 & .59 & -4.8 & 0.41& 5.7& 2.0& 3.4\\
$C$ & 5.9&$6.3\cdot 10^{4}$ & .023& .50 & 15 & 3.4 & -3.4 & -15& 3.1\\
  \hline
\end{tabular}
\end{center}
\caption{\label{fit}\it Fit results. Shown are the best fit values of the five independent parameters defining the 
two-zero textures as well as the obtained values of the mixing angles, the $\chi^2$ and the fine-tuning $d_{FT}$ of the minimum. Results refer to the NH case.} 
\end{table}
\noindent
First of all, we see from Tab.\ref{fit} that for all patterns but C we can find a very small $\chi^2$; 
this is not particularly surprising since, as claimed in the Introduction, all patterns are compatible with the 
experimental data and C prefers the inverted hierarchy. Differences arise as we deal with the chosen estimators, the magnitude and hierarchies of the matrix elements
and the values of the mixing parameters corresponding to the best fit points. In fact, 
patterns $A_{1,2}$ show good agreement of all observables with Tab.\ref{data}
and the fitted parameters (matrix elements) are close to $1$, with only modest $d_{FT}$ values.
On the contrary, although the textures $B_{i=1,...,4}$ also have small $\chi^2$'s,  the fine-tuning among the matrix elements is quite 
strong, as it can be seen from the larger $d_{FT}$ values compared with the $A_{1,2}$ ones. In addition, 
a more pronounced hierarchy between the parameters is needed
to fit the data. The $B_i$ textures only differ for the preferred $\theta_{23}$;
% cannot be treated on the same footing because $B_{1,3}$ present smaller
% fine-tuning and $\chi^2$ than $B_{2}$ and $B_{4}$; 
the reason can be traced back to eq.(\ref{p23}): $B_{1,3}$ prefer an atmospheric 
angle less than $\pi/4$ and very close to the best fit $\sin^2 \theta_{23}\sim 0.41$ whereas textures $B_{2,4}$ would rather 
favor a $\theta_{23} > \pi/4$, very close to the other best fit $\sin^2 \theta_{23}\sim 0.59$. Notice that these results are slightly dependent on the 
set of data used for the fit; if, instead of Tab.\ref{data}, we had adopted the best fits and errors as given in \cite{Schwetz:2011zk},
with only one minimum at $\sin^2\theta_{23}\sim 0.52$, the situation would have been different with, for example, a better $\chi^2$ and smaller $d_{FT}$ for 
$B_2$ compared to $B_1$. This would also be the case for $A_{1,2}$ but, since these textures do not prefer any octant,
both matrices perform very well.
The case $C$ is the more finetuned and it prefers $\theta_{23} = \pi/4$, that is partially excluded with the current data.

Summarizing, textures $B_i$ and $C$ can be classified as {\it less natural} with respect to $A_{1,2}$ because 
of the larger fine-tuning parameter $d_{FT}$ and the more pronounced hierarchies among the $a_i$ matrix elements.
It is important to stress that, given the relatively large number of fitted parameters, for some textures several good 
$\chi^2$ comparable in size can be found, with matrix elements slightly different from fit to fit. In these cases, we observed
that the smaller $\chi^2$ is usually given by very fine-tuned solutions with a strong hierarchy between the $a_i$ and, as a 
selecting criteria,  we decided to present  the results with the smaller value of the product $\chi^2\cdot d_{FT}$.

\section{Texture predictivity}
\label{textpre}
From the model building point of view, it could be interesting to describe the previous textures in terms of powers of the
Cabibbo angle $\lambda$. Taking the values of the $a_i$ coefficients in  Tab.\ref{fit} and 
%normalizing their ratios to the largest matrix elements, we get:
maintaining the hierarchies among them we get:
\begin{eqnarray}
\label{text1}
\tilde A^{}_1: ~~ \left(\matrix{ 0 & 0 & \lambda \,\tilde a_1 \cr 
0 &  \tilde a_2 & \tilde a_3 
 \cr \lambda  \,\tilde a_1 & \,\tilde a_3 &  \tilde a_4 \cr} \right) \, , ~~~~ 
\tilde A^{}_2: ~~ \left(\matrix{ 0 & \lambda\,\tilde a_1 & 0 \cr \lambda\,\tilde a_1 & \tilde a_2 &
\,\tilde a_3 \cr 0 & \tilde a_3 & \tilde a_4 \cr} \right) \, ;
%     (2)
\end{eqnarray}
%and
\begin{eqnarray}
\label{text2}
\tilde B^{}_1: ~~ \left(\matrix{ \tilde a_1 & \lambda^2\,\tilde a_2 & 0 \cr \lambda^2\,\tilde a_2 & 0 &
\tilde a_3 \cr 0 & \tilde a_3 & \lambda \,\tilde a_4\cr} \right) \, , \qquad 
 \tilde B^{}_2: ~~
\left(\matrix{ \tilde a_1 & 0 & \lambda^2\,\tilde a_2 \cr 0 & \lambda\,\tilde a_3 & \tilde a_4 \cr
\lambda^2\,\tilde a_2 & \tilde a_4 & 0 \cr} \right) \, , ~~~~~
\qquad
\tilde B^{}_3: ~~ \left(\matrix{\tilde a_1 & 0 & \lambda^2\,\tilde a_2 \cr 0 & 0 &
\tilde a_3 \cr \lambda^2 \,\tilde a_2& \tilde a_3 & \lambda\,\tilde a_4 \cr} \right)  \, 
%     (3)
\end{eqnarray}
%and
\begin{eqnarray}
\label{text3}
\tilde B^{}_4: ~~ \left(\matrix{ \tilde a_1 & \lambda^2\,\tilde a_2 & 0 \cr \lambda^2 \,\tilde a_2& \lambda\,\tilde a_4 &
\tilde a_3 \cr 0 & \tilde a_3 & 0 \cr} \right)\, , \qquad 
\tilde C: ~~ \left(\matrix{\tilde a_1 & \lambda\,\tilde a_2 & \lambda\,\tilde a_3 \cr \lambda\,\tilde a_2 & 0 &
\tilde a_4 \cr \lambda\,\tilde a_3 & \tilde a_4 & 0 \cr} \right) \; ,
%     (4)
\end{eqnarray}
where the new parameters $\tilde a_i$ are unspecified ${\cal O}(1)$ numbers. In the spirit of $U(1)$ constructions, the coefficients of every matrix elements are unspecified 
complex entries with absolute values of ${\cal O}(1)$; then, 
we have not expressly shown any possible correlation among the matrix elements, although they can be important to get 
the correct values of the neutrino mixing parameters. The only thing we can say is that there exists a non-vanishing volume in the
parameter space where the neutrino data can be fitted with sufficient accuracy.

\vskip .3 cm 
Some analytical considerations on the predicted masses and mixing angles are possible by means of standard perturbation theory in the 
expansion parameter $\lambda$ applied to eqs.(\ref{text1}-\ref{text3}). It turns out 
that $\tilde A_{1,2}$ predict almost maximal $\theta_{23}$ and a value of the reactor angle as large as $\lambda$; on the other hand, at leading 
order (LO) $\theta_{12}$ is vanishing and receives corrections at ${\cal O}(\lambda)$.
Textures of structure similar to $\tilde B_{1,2,3,4}$ have been carefully studied in \cite{Altarelli:2004za} where, however, some of their many properties 
have been obtained in the hypothesis that the matrix elements have well-definite values and not unspecified ${\cal O}(1)$ coefficients in
front of them. For vanishing $\lambda$, the $\tilde B_{1,2,3,4}$ have two degenerate eigenvalues; if the third one is the largest, then we can associate
the two degenerate masses to the solar sector and $(1,0,0)^T$ to $m_3$. When also the corrections from $\lambda$ are taken into account, the previous assignments
produce small $r=\Delta m^2_{sol}/\Delta m^2_{atm}$ and very large $\theta_{13}$. The solar and atmospheric angles are unstable and can get almost any value. On the other hand, 
if the non-degenerate mass is smaller than the degenerate pair of eigenvalues, then we are forced to assign such a smaller mass to $m_1$ and the other two 
eigenvalues to $m_2$ and $m_3$. Again, reintroducing $\lambda$, we get large $r$, small $\theta_{12,13} \sim {\cal O}(\lambda^2)$ and the correct value $\theta_{23} = \pi/4 +{\cal O}(\lambda)$.
This last case seems to agree better with the experimental results (as it can be seen from the values of the $a_i$ in Tab.\ref{fit}) although a moderate finetuning on the parameters is needed (see later for the correlations).
Texture $\tilde C$ has the same LO as for $\tilde B$ so, again, we have to distinguish two cases as before. 
When the almost degenerate eigenvalues are associated to $m_{1,2}$, then $r$ is vanishingly small, $\theta_{13,23}\sim {\cal} O(1)$ and the solar 
angle is always small, suppressed by the (11) entry of the neutrino mass matrix. In the opposite case, $r\sim 1$, 
$\theta_{13,12}\sim {\cal} O(\lambda)$ and $\theta_{23} \sim \pi/4$.
Again this last case is favoured by the data (always at the price of a large finetuned correlations between the $a_i$).

%Based on these rough estimates, we see that the most problematic angle is $\theta_{12}$ and that $r$ is mostly close to unity: both can only be reproduced accidentally by a conspiracy of the ${\cal O}(1)$ coefficients. On the other hand, there is no difficulty in obtaining a naturally large atmospheric angle whereas $\theta_{13}$ is generally of ${\cal O}(\lambda)$, as for $\tilde A_{1,2},\tilde C$, or a bit smaller for the textures of type $\tilde B_i$.

\begin{figure}[h!]
\epsfig{file=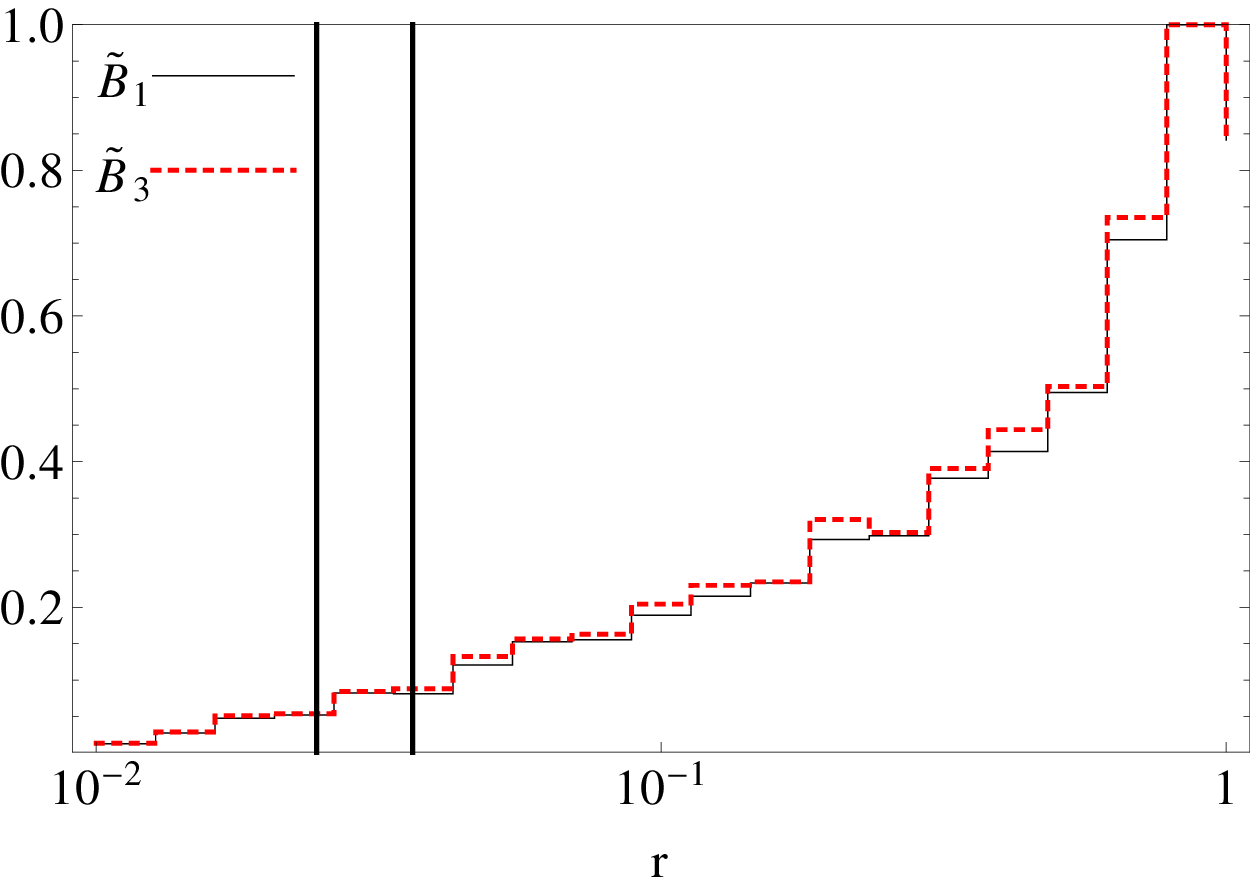,width=8cm}\qquad\epsfig{file=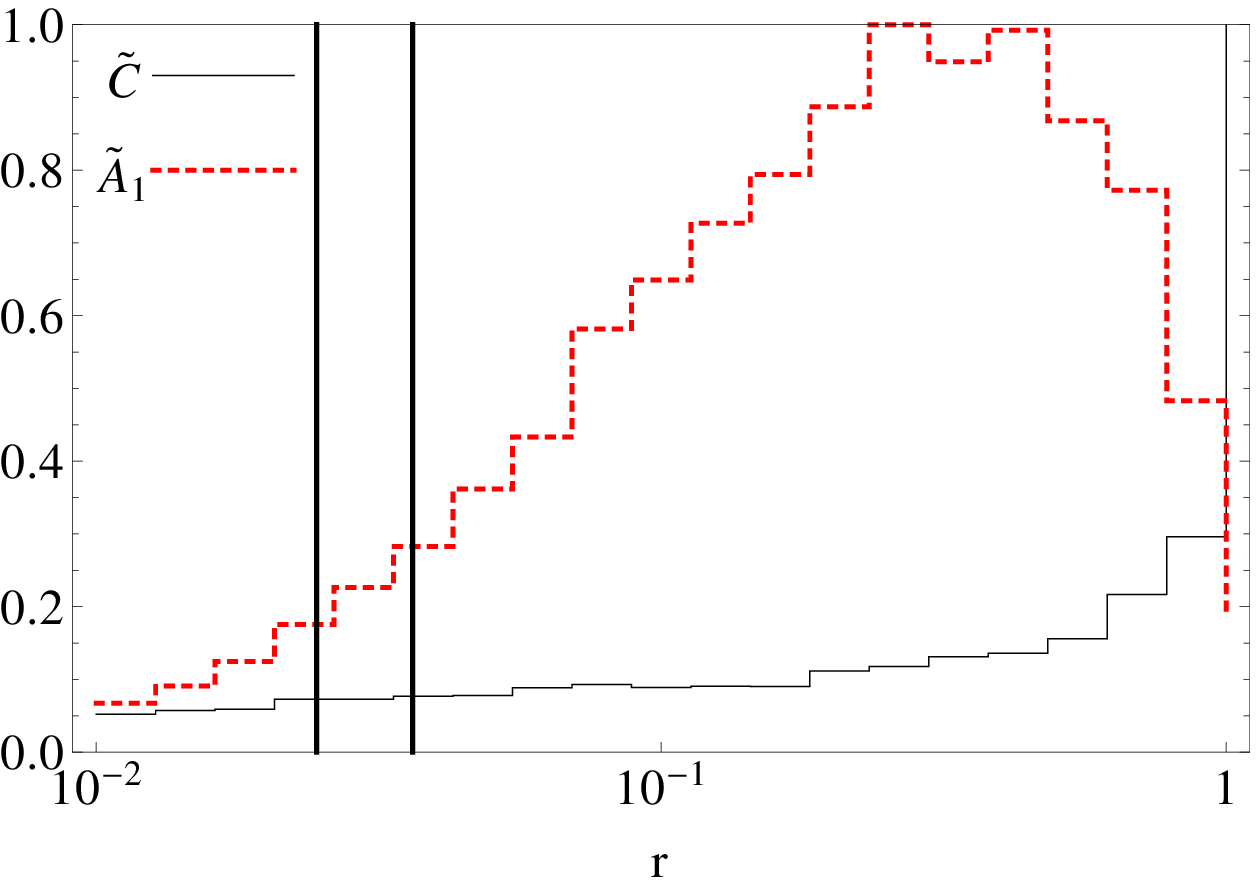,width=8cm}
\caption{\it \label{erre} Distribution for the variable $r$ as obtained from the textures  $\tilde B_1$ and $\tilde B_3$ (left panel - solid and dashed lines, respectively) 
and $\tilde A_1$ and $\tilde C$ (right panel - dashed and solid lines, respectively). The vertical black lines indicate the 3$\sigma$ allowed region on $r$.}
\end{figure}

\begin{figure}[h!]
\epsfig{file=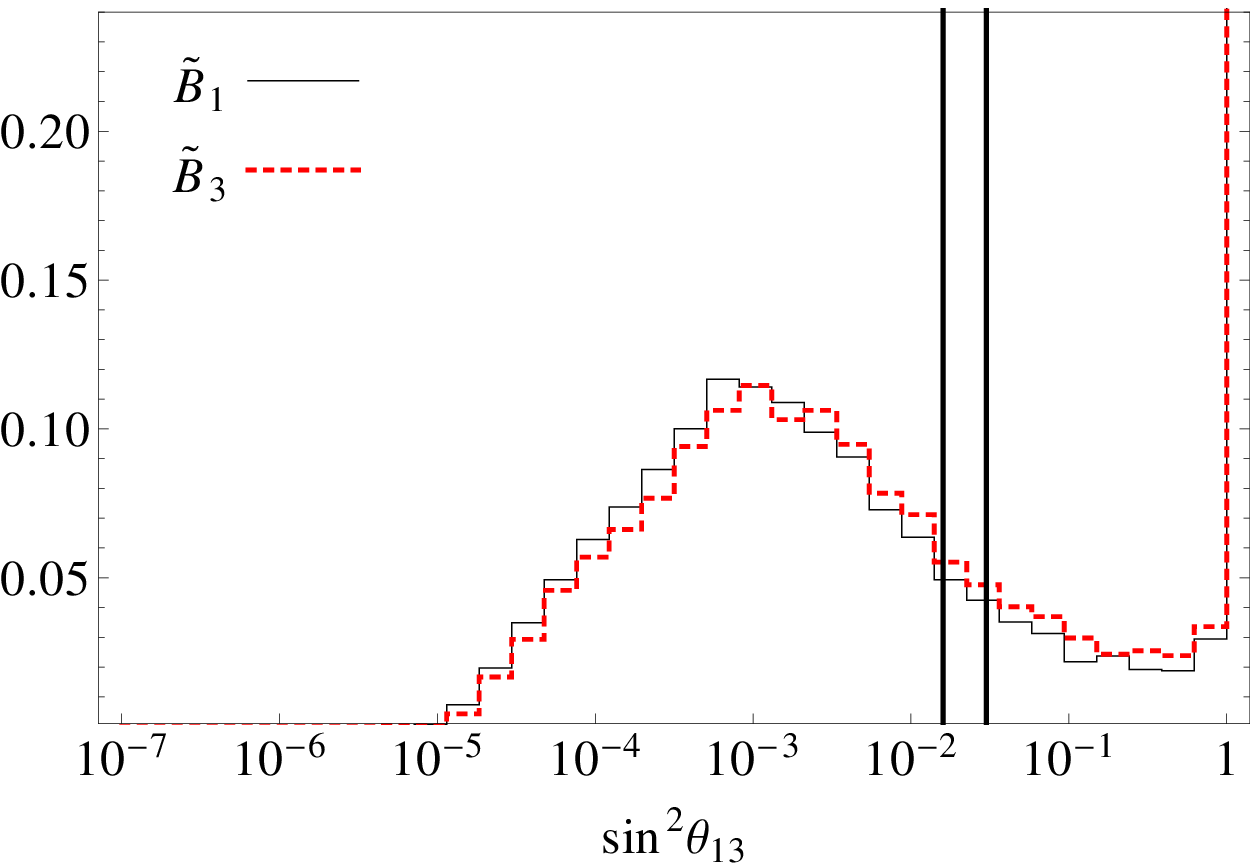,width=8cm}\qquad\epsfig{file=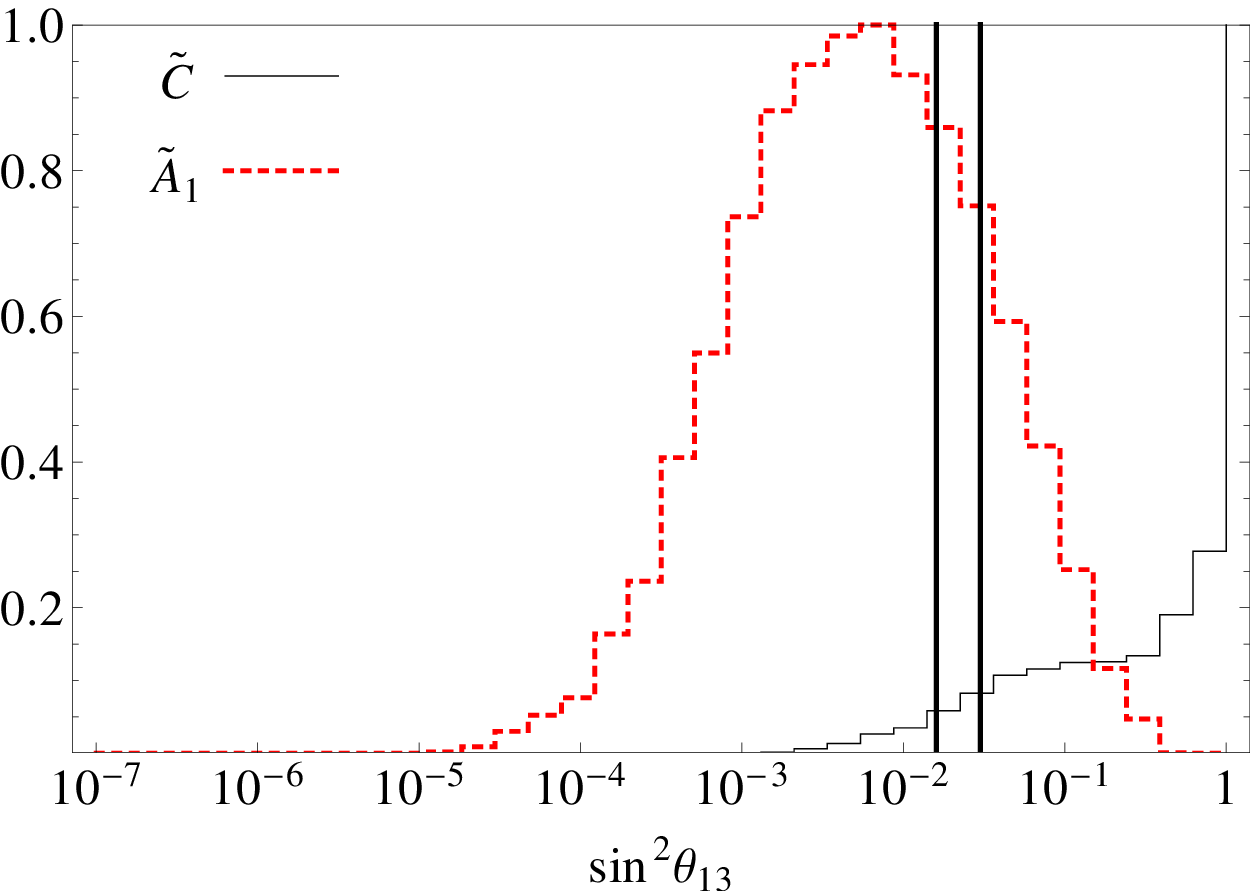,width=8cm}
\caption{\it \label{13} Distribution for the variable $\sin^2 \theta_{13}$ as obtained from the textures  $\tilde B_1$ and $\tilde B_3$ (left panel - solid and dashed lines, respectively) and
$\tilde A_1$ and $\tilde C$ (right panel - dashed and solid lines, respectively). The vertical black lines indicate the 3$\sigma$ allowed region on $\theta_{13}$.}
\end{figure}

\begin{figure}[h!]
\epsfig{file=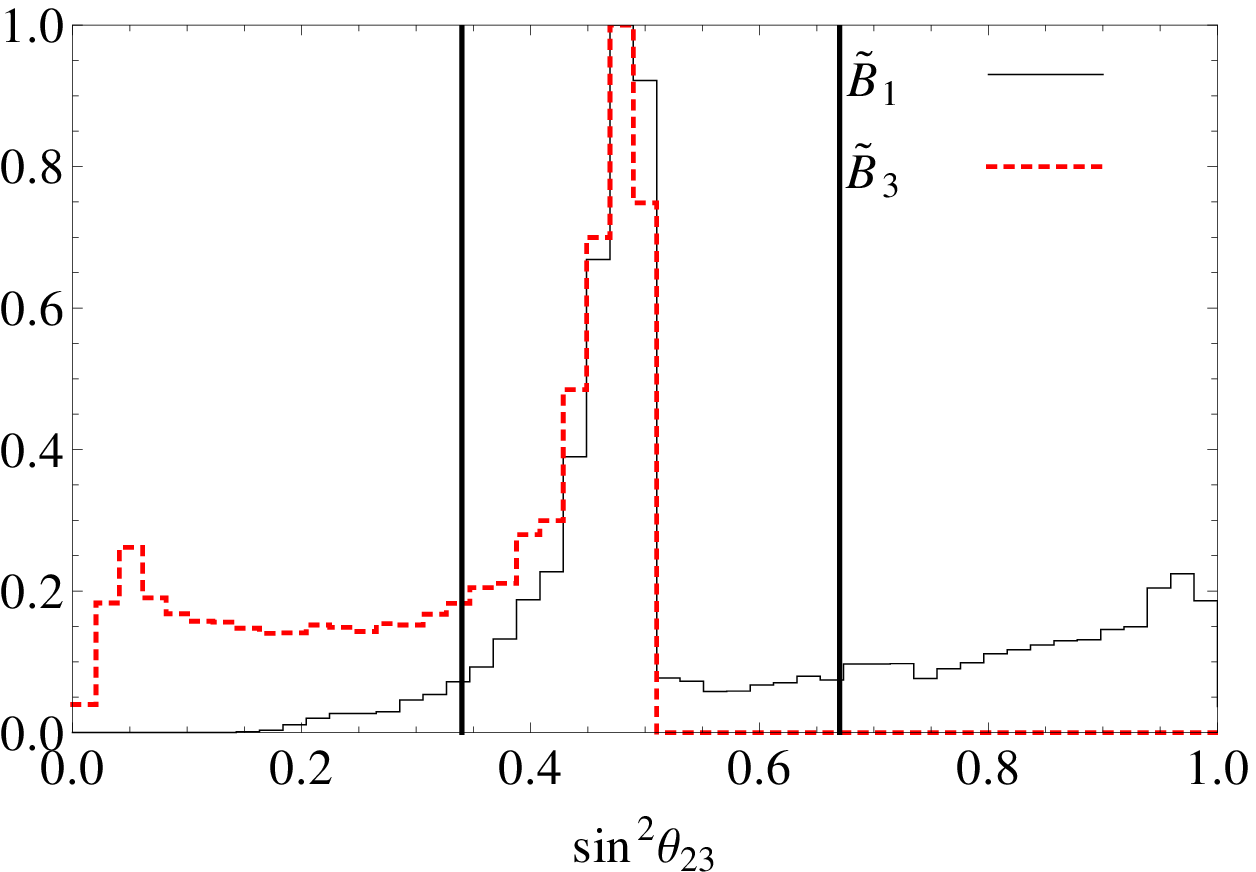,width=8cm}\qquad\epsfig{file=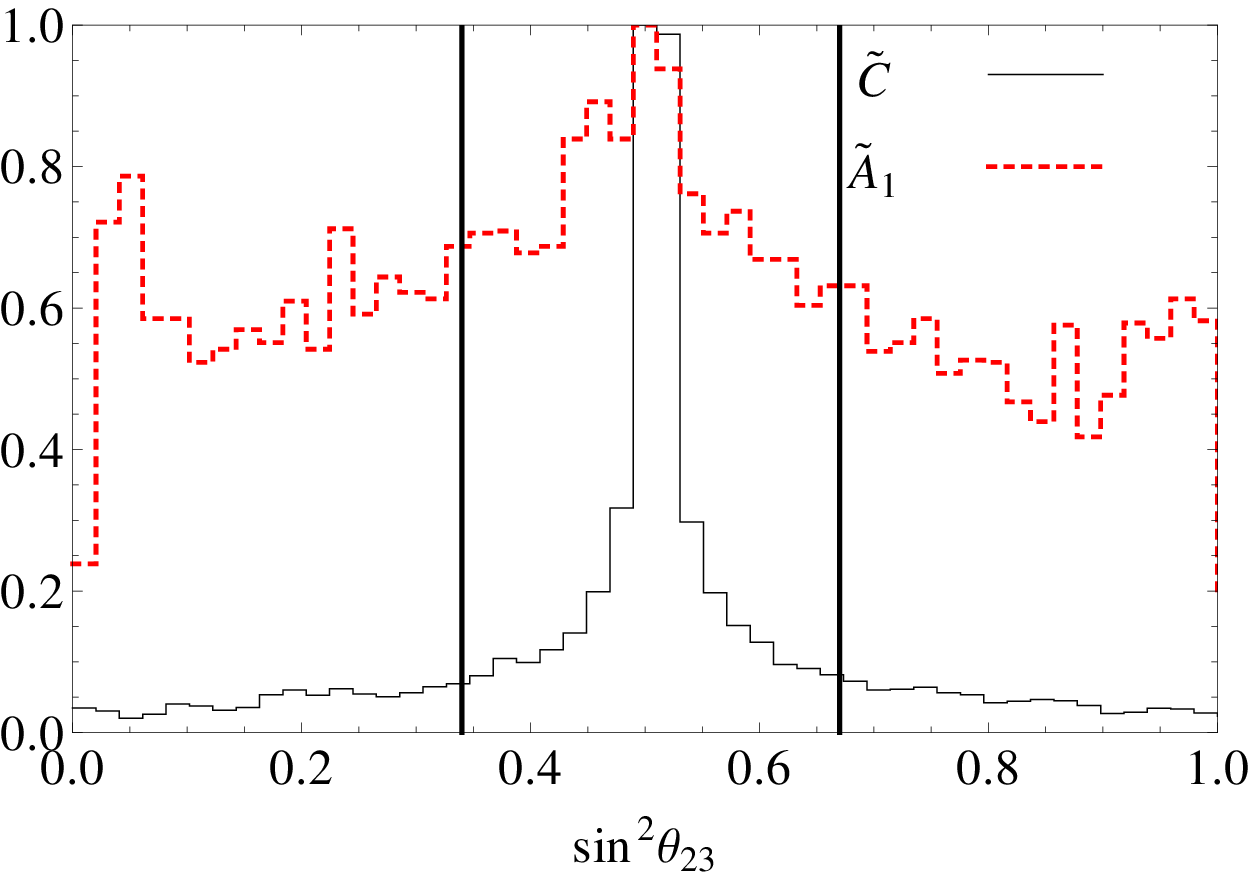,width=8cm}
\caption{\it \label{23}  Distribution for the variable $\sin^2 \theta_{23}$ as obtained from the textures  $\tilde B_1$ and $\tilde B_3$ (left panel - solid and dashed lines, respectively) and
$\tilde A_1$ and $\tilde C$ (right panel - dashed and solid lines, respectively). The vertical black lines indicate the 3$\sigma$ allowed region on $\theta_{23}$.}
\end{figure}

\begin{figure}[h!]
\epsfig{file=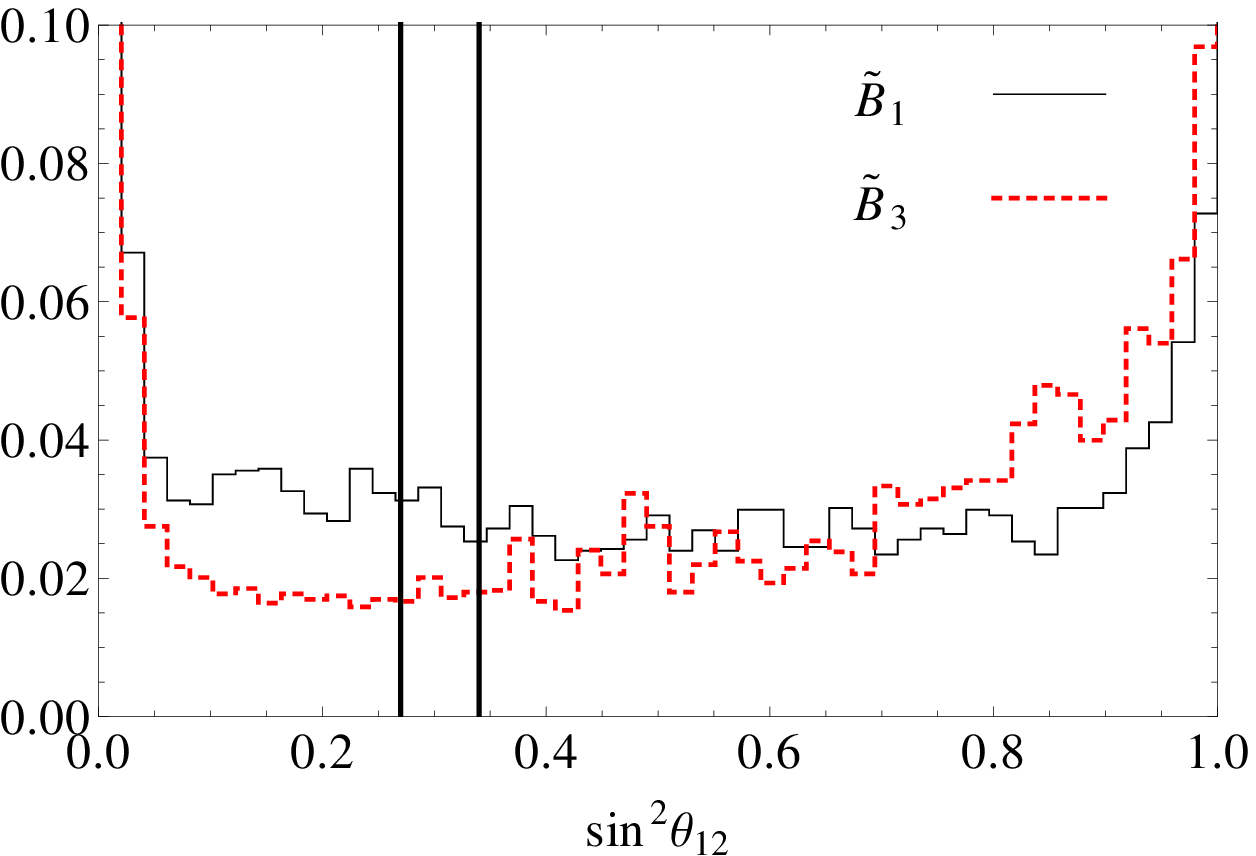,width=8cm}\qquad\epsfig{file=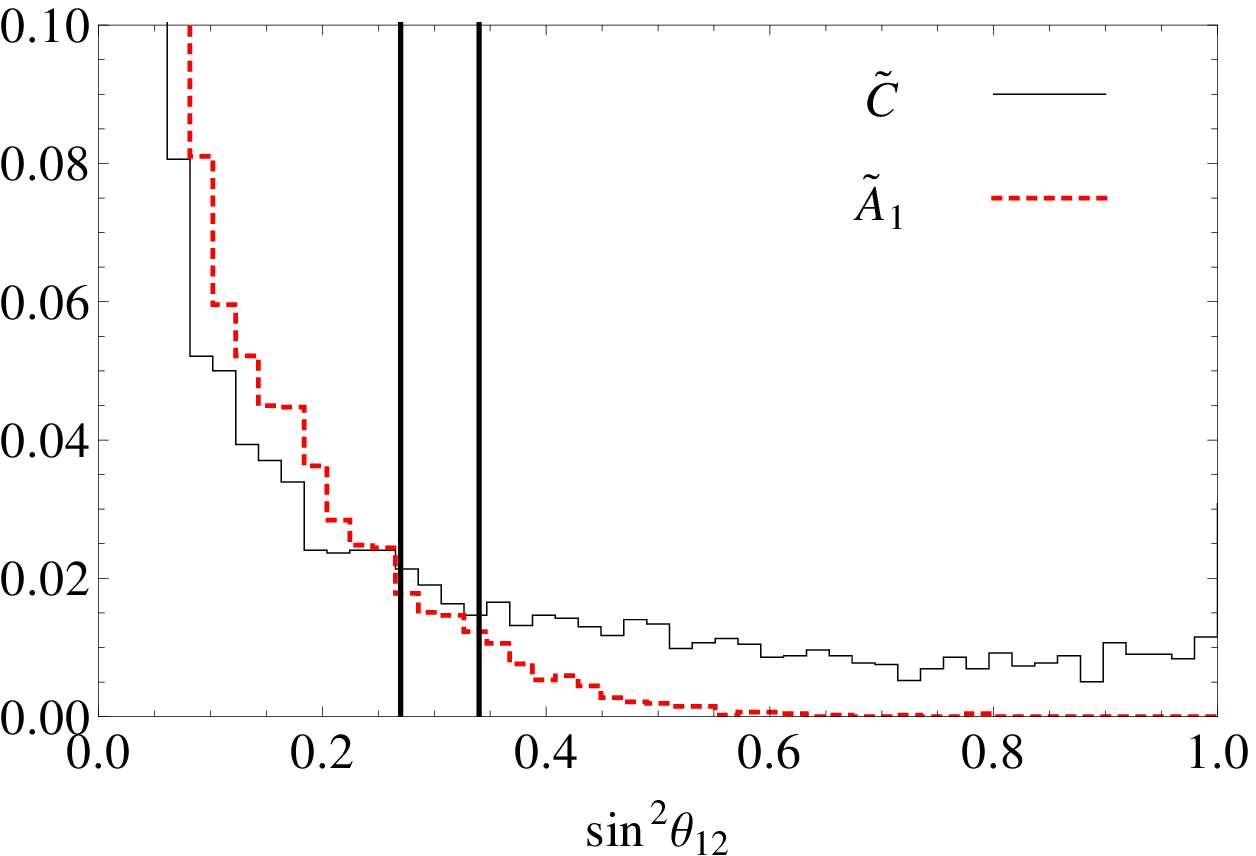,width=8cm}
\caption{\it \label{12}  Distribution for the variable $\sin^2 theta_{12}$ as obtained from the textures  $\tilde B_1$ and $\tilde B_3$ (left panel - solid and dashed lines, respectively) and
$\tilde A_1$ and $\tilde C$ (right panel - dashed and solid lines, respectively). The vertical black lines indicate the 3$\sigma$ allowed region on $\theta_{12}$.}
\end{figure}

\vskip .3 cm 
These considerations are confirmed by a numerical analysis; we performed a MonteCarlo simulation
extracting randomly all {\cal O}(1) entries of eqs.(\ref{text1}-\ref{text3})
with moduli in $[1/3,3]$ (with equal probability of obtaining a number smaller than 1 and larger than 1);
the unique phase $\varphi$ is extracted flat in $[-\pi,\pi]$ and can always be associated to $\tilde a_4$
\footnote{We choose $a_4$ just for convenience. In fact, it can be shown analitically 
that, using the freedom in the definition of the neutrino fields, the physical phase can accompain any matrix elements but 
the one in a row (or column) with two vanishing entries. 
% Among the former, then, the phase can be placed  
% everywhere and this does not have any impact on the distributions of the physical neutrino parameters. 
% If a model builder gets a mass matrix of the type described in our paper with the phase $\phi$
% not associated to $a_4$, then $\phi$ can be moved to $a_4$ by appropriately redefining the neutrino fields.
The position of the phase does not affects the fine-tuning either; in fact, the physical observables are complicated functions of the 
matrix elements of ${\cal O}(1)$ and $\varphi$ and, given the large number of terms, a  value of an observable
can be obtained slightly changing the values of the $a_i$ and $\varphi$, with no need to 
force or modify any correlations.}; 
we then built the hermitian matrix $m_\nu^\dagger m_\nu$
and extracted the eigenvalues and eigenvectors. 
%Since we are focusing here on the normal hierarchy case, we order the eigenvectors in such a way the last column of $U_{PMNS}$ matrix corresponds to the largest eigenvalue and the first one to the smallest.
We do not impose any external additional constraints on the mass differences and mixing angles.
Notice that this is different from what studied, for example, in \cite{Fritzsch:2011qv} since we started from explicit 
mass textures (derived from our fit procedure) and use a typical $U(1)$ approach to estimate the predicted angles whereas the authors of \cite{Fritzsch:2011qv} generate sets of 
random numbers of experimentally allowed $(\theta_{12},\theta_{13},\theta_{23})$ and mass differences to predict other quantities of interest and retaining those 
mixing parameters satisfying certain consistency relations.
We plot the distributions of the ratio $r=\Delta m^2_{sol}/\Delta m^2_{atm}$ and $\sin^2 \theta_{13}$, $\sin^2 \theta_{23}$ and 
$\sin^2 \theta_{12}$ in Figs.\ref{erre},\ref{13},\ref{23} and \ref{12}, respectively. 
Given the $23$-symmetry discussed around eq.(\ref{p23}), which relates the mixing angles among those textures connected by $P_{23}$, 
we do not show the obtained distributions for all textures but rather prefer to superimpose in the left panel the 
distributions for the textures $\tilde B_1$ and $\tilde B_3$ (solid and dashed lines, respectively) and in the right panel the results for $\tilde A_1$ and $\tilde C$ 
(dashed and solid lines, respectively).
The vertical black lines indicate the 3$\sigma$ allowed region on the variable under discussion.

\vskip .3 cm 
As expected, it is difficult to reproduce the small value of $r\sim 0.03$ in a natural way, since it usually requires a large amount of fine-tuning
on the mass parameters. We observed only a moderate depletion of $r$ for the patter $\tilde A_{1}$, $r \sim 0.5$ (Fig.\ref{erre}).
From Fig.\ref{13} we see that $\tilde A_1$ is very powerful in reproducing the relatively 
large values of $\theta_{13}$, a feature not shared by the textures $\tilde B_{1,3}$ since they
present a peak in the distribution at the lower bound of the 3$\sigma$ 
range.
Textures $\tilde C$ performs worse than the others, since even the smallest peak at $\sin^2 \theta_{13}\sim 0.1$ is essentially excluded.  We take this as an indication 
that the texture $\tilde C$ is not particularly suitable to fit the data in  normal hierarchy.
The angle $\theta_{23}$ (Fig.\ref{23}) is well reproduced by essentially all textures, with deviations as discussed below eq.(\ref{text3}). 
Also, the distribution of $\theta_{12}$ is almost flat for  $\tilde B_{1,3}$ and favors 
small values for $\tilde A_1$ and $\tilde C$ (Fig.\ref{12}).

\vskip .3 cm 
It is interesting to evaluate the fraction of points which simultaneously give $\theta_{ij}$ and $r$ into their the 3$\sigma$ bounds.
This can be considered as another estimator of the fine-tuning among the matrix elements needed to reproduce the correct experimental data. 
The computation results in the following small fractions:
\begin{eqnarray}
\{\tilde A_1,\tilde A_2,\tilde B_1,\tilde B_2,\tilde B_3,\tilde B_4,\tilde C\} = 
\{0.1,0.1,0.004,0.003,0.003,0.002,<0.0001
\}\%\,,
\end{eqnarray}
confirming that the textures $\tilde A_{1,2}$, derived from $A_{1,2}$ of eq.(\ref{aini}),
are the most appropriate to describe the whole data in the neutrino sector in NH.

\begin{figure}[h!]
\hspace*{-1.5cm}
\epsfig{file=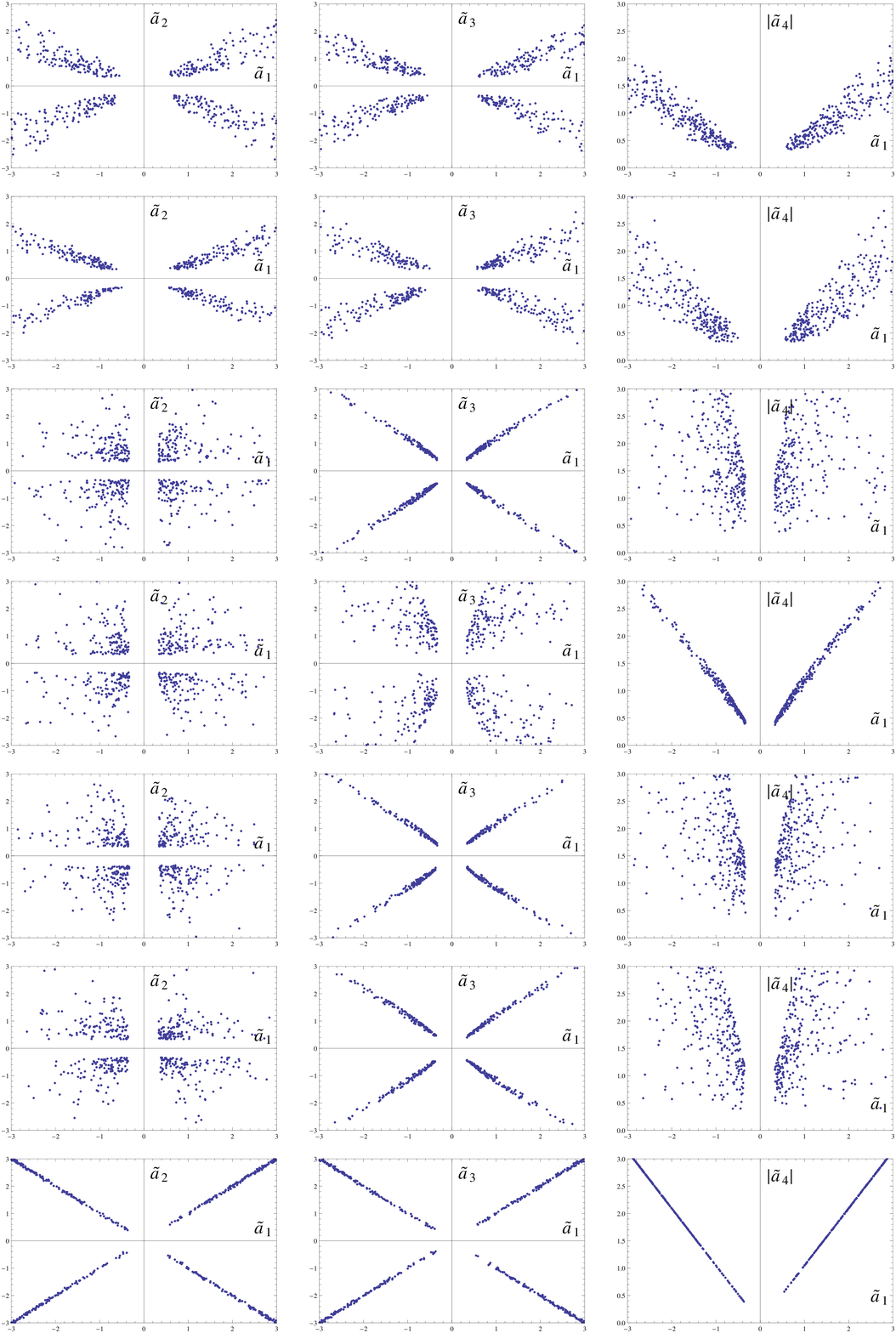,width=15cm}
\caption{\it \label{corrNH} Correlations among $\tilde a_1$ and 
$\tilde a_{2,3}, |\tilde a_4|$ for the textures studied in the paper, for the case of NH. From above to below:
$\tilde A_1,\tilde A_2,\tilde B_1,\tilde B_2,\tilde B_3,\tilde B_4,\tilde C$.}
\end{figure}

\vskip .3 cm 
The previous plots do not give any indications of the degree of correlations among the matrix elements $\tilde a_i$ of 
eqs.(\ref{text1}-\ref{text3}) needed to get the desired values of the mixings and $r$. 
This is illustrated in Fig.\ref{corrNH}, where we show, for all seven textures, the correlations among $\tilde a_1$ and 
$\tilde a_{2,3}, |\tilde a_4|$.

The results of Fig.\ref{corrNH} show that the textures $\tilde{A}_i$ reproduce the data at the price of mild correlations between all the parameters,
whereas we always found a significant correlation between a pair of $\tilde{a}_i$ (the ones corresponding to the largest in Tab.\ref{fit}) for $\tilde{B}_i$.
The strong correlation observed for the $\tilde C$ is a clear signal that such a texture is specially suitable to describe 
the data in IH more than in NH, so the matrix elements have to conspire in order to reproduce the data in the "wrong" hierarchy.
%We notice that, dividing the points by the appropriate power of $\lambda$ for those entries in eqs.(\ref{text1}-\ref{text3}) whose $\lambda$ factors have been stripped from the matrix elements, we automatically get also the correlations among the matrix elements of the original textures.

\vskip .3 cm 
We also observe that the degree of fine tuning for the $\widetilde{A_{i}}$, $\widetilde{B_{i}}$ and $\widetilde{C}$ textures is the same 
as for the original matrices because numerator and denominator in the definition of $d_{FT}$  scale
with the same power of $\lambda$.

\section{Results for the inverted hierarchy}
\label{inverted}

We have repeated the same fit procedure and analysis for the inverted hierarchy (IH). The main results are:
\begin{itemize}
 \item 
the textures $A_{1,2}$ have now a very worse $\chi^2_{min}\sim {\cal O}(100)$ and then 
are not suitable to describe the neutrino data; we do not discuss such cases any further; 
\item the agreement of the texture $C$ is improved with respect to the normal hierarchy case, both in terms of fine-tuning 
and correlations; the corresponding expression expanded in terms of $\lambda$, $\tilde C^{IH}$ is given by:
\begin{eqnarray}
\label{text3IH}
\tilde C^{IH}: ~~ \left(\matrix{ 1 & 1 & 1 \cr 1 & 0 &
1 \cr 1 & 1 & 0 \cr} \right) \; ;
%     (4)
\end{eqnarray}
\item the main features of the textures $B_{i}$ remain essentially the same, with large correlations among the parameters, so that the expanded textures are the same as for the normal hierarchy case, that is: $\tilde B^{IH}_{1,2,3,4}=\tilde B_{1,2,3,4}$; 
\end{itemize}
For the sake of completeness we report such results in Tab.\ref{fitIH}.
\begin{table}[h!]
\begin{center}
\begin{tabular}{cccccccccc}
  \hline
  \hline
  texture & $\chi^2_{min}$ &$d_{FT}$ & $\sin^2 \theta_{13}$
  &  $\sin^2\theta_{23}$ & $a_1$ & $a_2$ & $a_3$ & $a_4$ & $\varphi$\\
  \hline
$B_1$ & $1.8\cdot 10^{-7}$ &$9.3\cdot 10^{3}$& .023 & .59 & -6.7 & 0.42 & -5.6  & -2.1  & 3.3 \\
$B_2$ & $4.4\cdot 10^{-7}$&$9.3\cdot 10^{3}$& .024 & .41 & -6.7  &-0.42 & -2.1 & 5.6  & -1.7\\
$B_3$ & $1.0\cdot 10^{-6}$& $9.2\cdot 10^3$ & .023 & .59 & 6.9& 0.48 & 5.9 & -2.0 &-0.14\\
$B_4$ &$3.1\cdot 10^{-7}$& $9.2\cdot 10^3$  & .023 & .41  & 6.9 & 0.48  & 5.9  &-2.0  & 0.13\\
$C$ & $2.9\cdot 10^{-7}$&$4.8\cdot 10^3$&  .023&  .59  &4.1 & -3.2 & -3.8 & -4.2 & -0.014\\
  \hline
\end{tabular}
\end{center}
\caption{\label{fitIH}\it Fit results for the inverted hierarchy. Shown are the best fit values of the five independent parameters defining the 
two-zero textures as well as the obtained values of the mixing angles, the $\chi^2$ and the fine-tuning $d_{FT}$ of the minimum.}
\end{table}
% The main difference with respect to the NH case is that the fine-tuning $d_{FT}$ is less pronounced for the textures $B_{2,4}$ than for $B_{1,3}$. 
% This is due to the fact that $B_{1,3}$ prefer $\theta_{23}>\pi/4$ and so, for the $P_{23}$ symmetry, $B_{2,4}$ prefer the best fit region.
% 

\vskip .3 cm 
The analytical discussion for the textures $\tilde B^{IH}$  can proceeds along the same lines as for the normal hierarchy case: in fact, 
the smallest mass $m_3$ can be associated to the eigenvector $(1,0,0)^T$ (thus giving small $r$ and close to maximal $\theta_{13}$, with
undetermined $\theta_{12,23}$) or to $(0,1,-1)^T$ (producing $r\sim {\cal O}(1)$, vanishing $\theta_{12,13}$ and maximal $\theta_{23}$).
For the texture $\tilde C^{IH}$ we can get an analytical understanding of the predicted masses and mixing angles working 
with real coefficients close to $1$; in this case, the texture predicts (in absence of accidental cancellations) 
large mixings for every angles $\theta_{13,12,23}$ and quite a large mass ratio $r$. Except for the atmospheric angle, 
whose values are broadened in the whole $[0,\pi/2]$ interval, these features are maintained as we allow for generic complex coefficients with 
absolute values in the $[1/3,3]$.

\begin{figure}[h!]
\epsfig{file=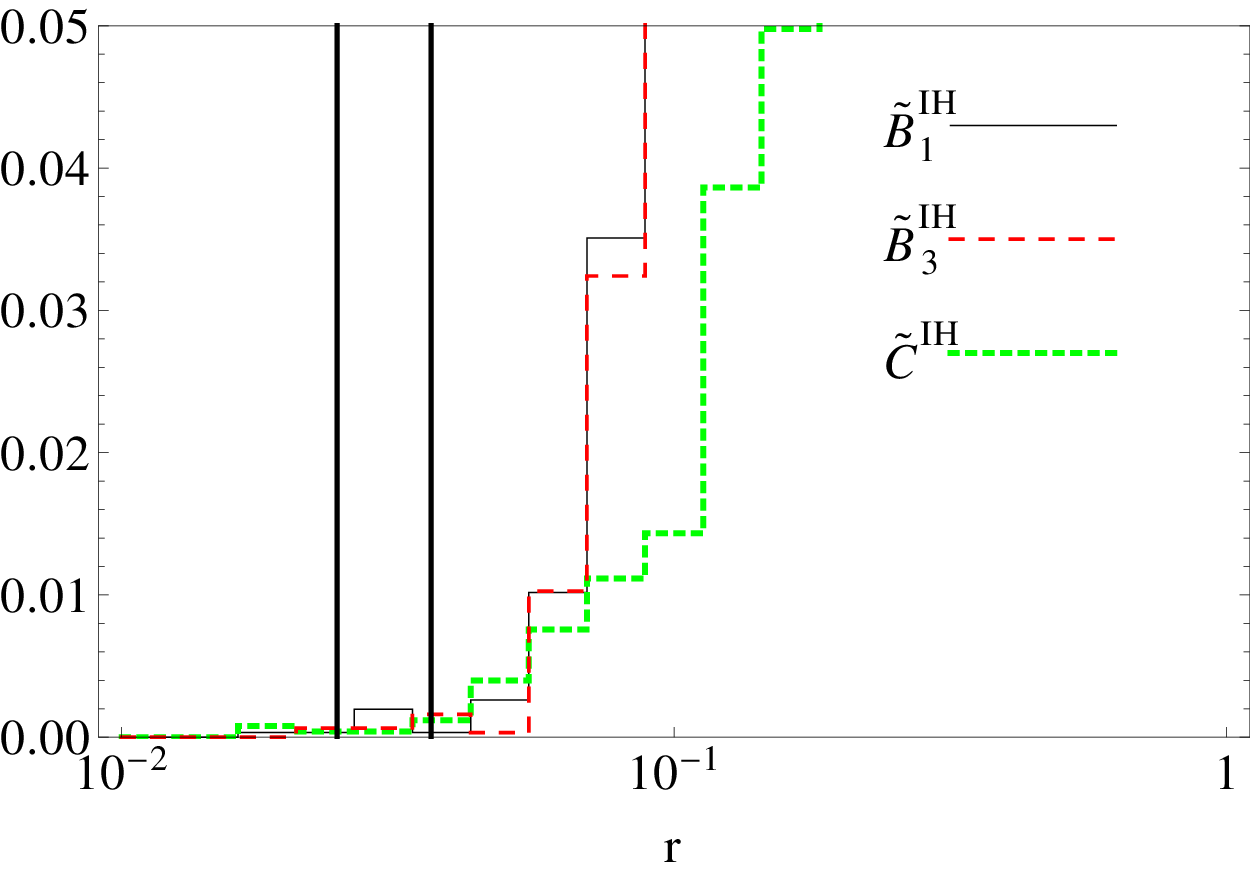,width=8cm}\qquad\epsfig{file=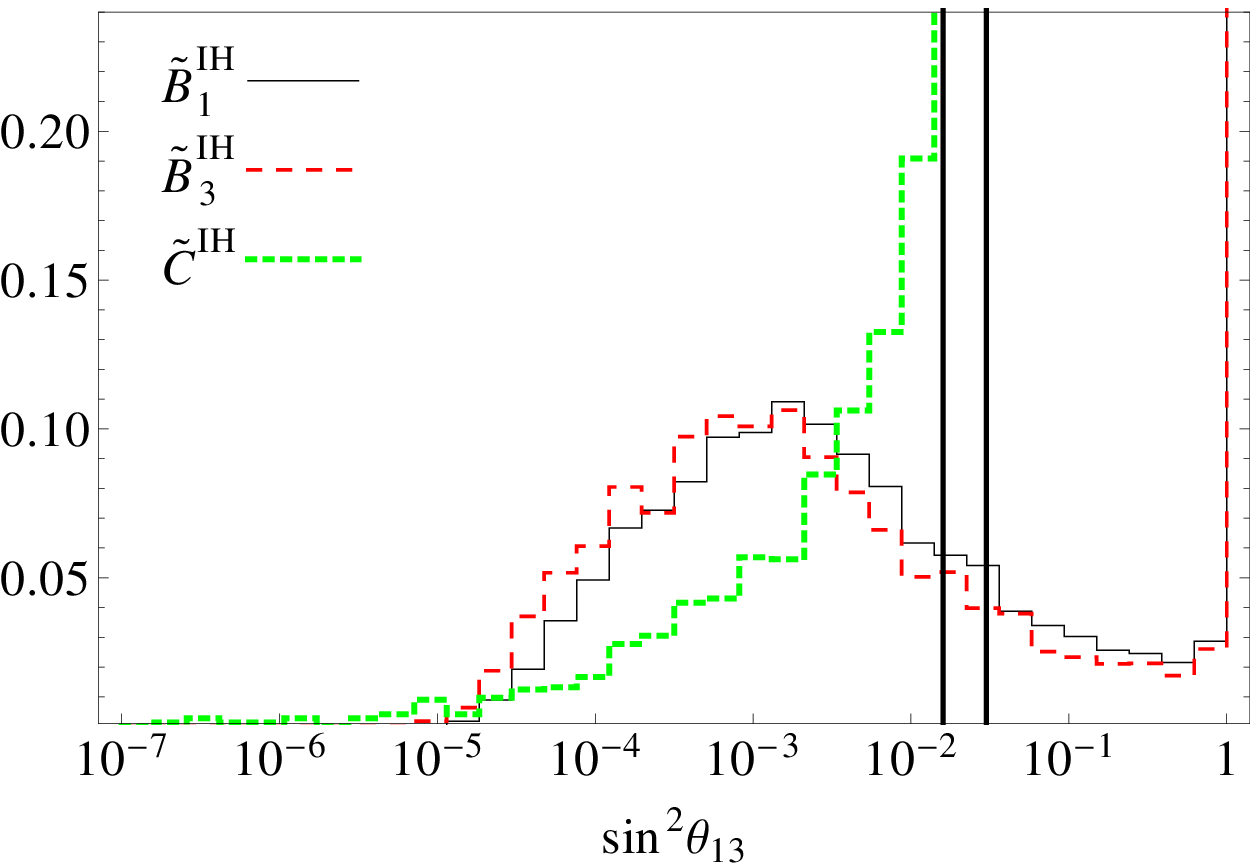,width=8cm}
\caption{\it \label{erre_IH} Distribution for the variable $r$ (left panel) and $\sin^2 \theta_{13}$ (right panel) as obtained from the textures 
$\tilde B^{IH}_1$ (solid line), $\tilde B^{IH}_3$ (dashed line) and $\tilde C^{IH}$ (thick dashed line). 
The vertical black lines indicate the 3$\sigma$ allowed region on $r$ and $\sin^2 \theta_{13}$.}
\end{figure}

\begin{figure}[h!]
\epsfig{file=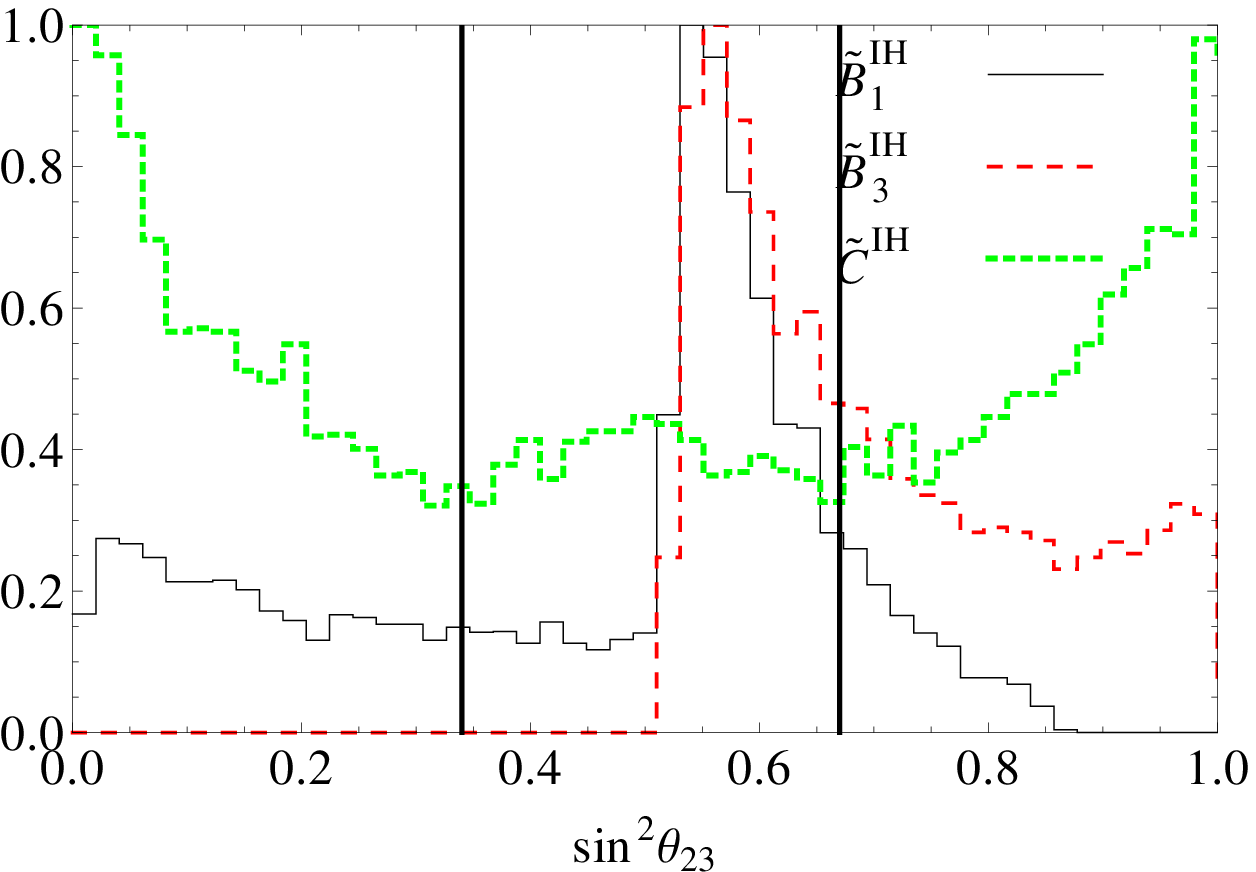,width=8cm}\qquad\epsfig{file=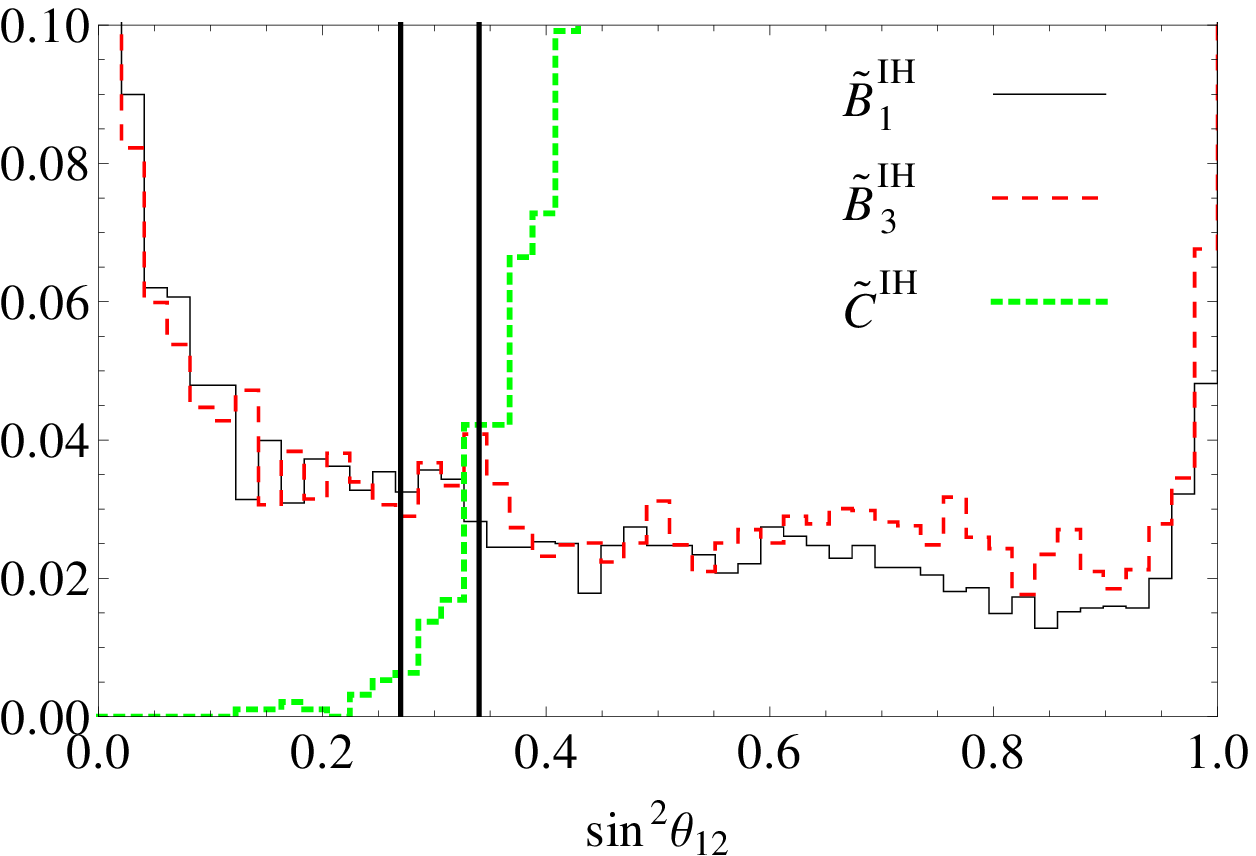,width=8cm}
\caption{\it \label{23_IH}  Distribution for the variable $\sin^2 \theta_{23}$ (left panel) and $\sin^2 \theta_{23}$ (right panel) as obtained from the textures 
$\tilde B^{IH}_1$ (solid line), $\tilde B^{IH}_3$ (dashed line) and $\tilde C^{IH}$ (thick dashed line). The vertical black lines indicate the 3$\sigma$ allowed 
region on $\sin^2\theta_{23}$ and $\sin^2\theta_{12}$.}
\end{figure}

\vskip .3 cm 
In Fig.\ref{erre_IH} and Fig.\ref{23_IH} we have shown the distributions resulting from the numerical analysis in which the matrix elements 
of the tilded textures  are extracted randomly with 
moduli in $[1/3,3]$ and phases in $[-\pi,\pi]$. Results are shown only for $\tilde B^{IH}_1$ (solid line), 
$\tilde B^{IH}_3$ (dashed line) and $\tilde C^{IH}$ (thick dashed line).
As it is common for inverted hierarchy, it is more difficult to reproduce the correct value of $r$ since this needs a stronger conspiracy 
among the matrix elements. This is the reason why the distributions shown in the left panel of Fig.\ref{erre_IH} are more 
shifted toward $r\sim 1$ than in the NH case but for $\tilde C^{IH}$, which gives very large $\theta_{13}$.
For the distributions of the other mixing angles, we do not observe any significant difference with respect to the NH case.

Summarizing, there is no preferred patterns in better agreement with the data: the texture $\tilde B_i^{IH}$ performs relatively
well for $\theta_{23}$ while many points obtained from $\tilde C^{IH}$ fall in the 3$\sigma$ range of $\theta_{13}$, and in every case very few realizations 
give the correct $r$ 
(much less then in the normal hierarchy case). 

\vskip .3 cm 
When we include the requirement for the three mixing angles and $r$ to be simultaneously into their 3$\sigma$ confidence levels, 
the fractions of successful trials is almost the same:
\begin{eqnarray}
\{\tilde B_1^{IH},\tilde B_2^{IH},\tilde B_3^{IH},\tilde B_4^{IH},\tilde C^{IH}\} = 
\{0.0008,0.0007,0.0008,0.0008,0.001
\}\%\,.
\end{eqnarray}

\vskip .3 cm 
As for the NH, we investigated the correlations among $\tilde a_1$ and $\tilde a_{2,3}, |\tilde a_4|$ for the textures 
$\tilde B_i$ and $\tilde C$ (given the large $\chi^2_{min}$, we do not need to study the $\tilde A_i$'s). 
\begin{figure}[h!]
\epsfig{file=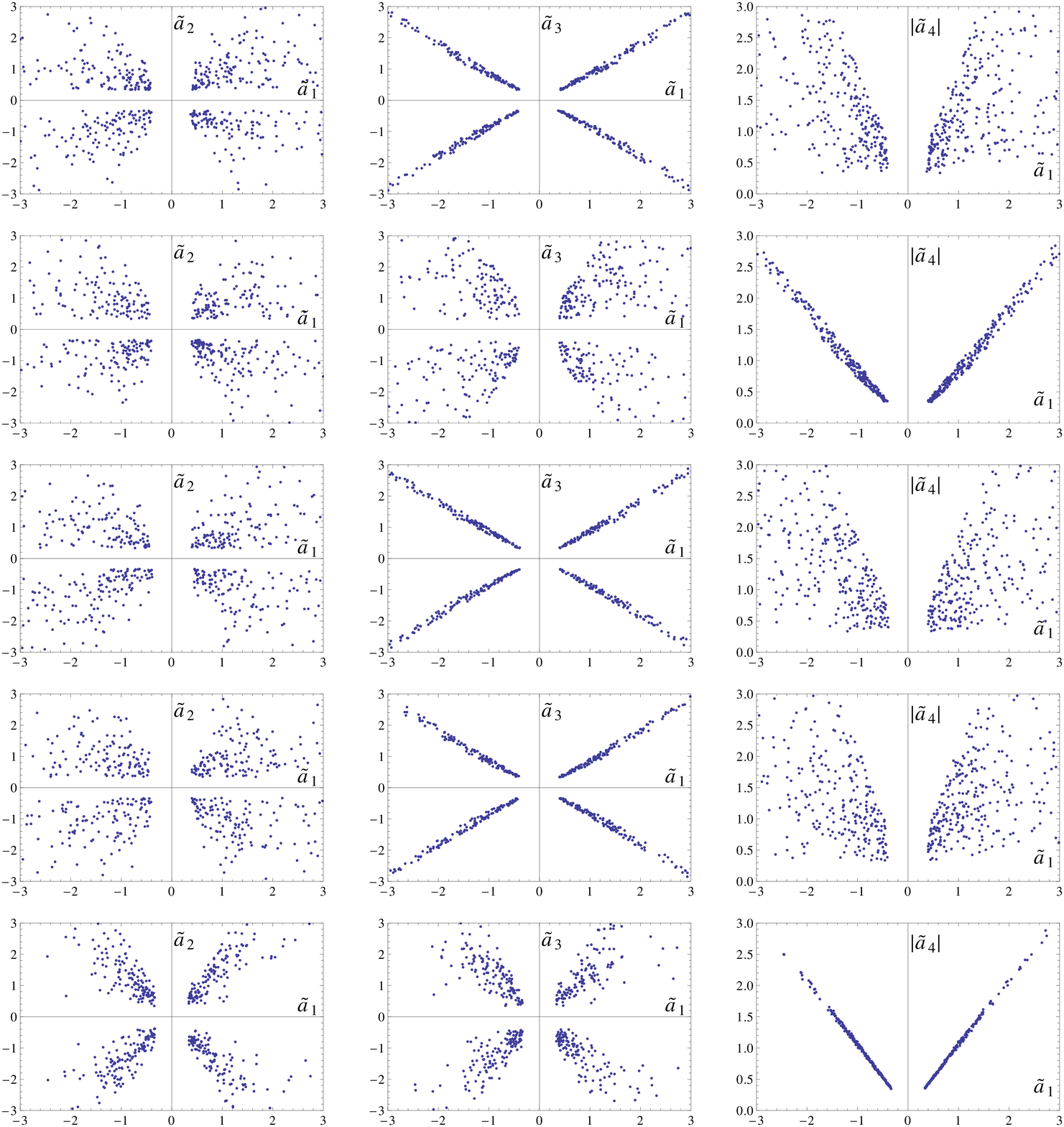,width=15cm}
\caption{\it \label{corrIH} Correlations among $\tilde a_1$ and 
$\tilde a_{2,3}, |\tilde a_4|$ for the textures studied in the paper, for the case of IH. From above to below:
$\tilde A_1,\tilde A_2,\tilde B_1,\tilde B_2,\tilde B_3,\tilde B_4,\tilde C$.}
\end{figure}
The qualitative behavior of the correlations for the $\tilde B^{IH}_i$ textures follows that of the NH case since, as explained in Sect.\ref{text}, 
the $B_i$'s describe equally well the NH and IH cases depending on the octant of $\theta_{23}$. As expected from those same 
considerations, the $\tilde a_i$ parameters of the 
$\tilde C^{IH}$ are now less correlated than in the NH case, since the $C$ mass matrix is more suitable to describe the inverted ordering 
of the mass eigenstates.

\section{Conclusions}
\label{concl}
In this paper we have studied the compatibility of certain type of two-zero Majorana neutrino mass textures with the recent data 
in the neutrino sector. Differently from what has been extensively studied in the literature, we used the 
neutrino data to get information directly on the parameters entering the mass textures, with the aim of understanding 
how naturally a given texture is able to reproduce the value of the mixing angles and mass differences. To achieve this aim,
we have used two different estimators: the minimum of the $\chi^2$, built to extract the values of the matrix elements which 
better reproduce the experimental data, and  the variable $d_{FT}$, defined in eq.(\ref{fine-tuning}), which is essentially a measure
of the fine-tuning among the parameters. 

We performed such an analysis for both normal and inverted ordering of the neutrino 
masses. In the first case, we observed that the patterns $B_i$ and $A_i$ fit the data with a very good $\chi^2$, although
the textures $A_{1,2}$ seem to be more favored. In fact patterns $A_i$ also show  a  relatively small $d_{FT}$ and a less pronounced hierarchy 
among the matrix elements, features not shared by the patterns $B_i$.
Texture $C$ gives the largest $d_{FT}$ and in addiction it needs $\theta_{23}$ very close to the maximal value, that currently seems disfavored.
Summarizing, textures $B_i$ and $C$ can be classified as {\it less natural} with respect to $A_i$ because 
of the larger fine-tuning parameter $d_{FT}$ and hierarchies among the matrix elements. 

Taking the central values of the obtained $a_i$'s, we have derived approximate
expressions of the textures in terms of the Cabibbo angle; extracting random ${\cal O}(1)$ coefficients, we have seen that such textures 
show a similar agreement with the data as for the original patterns. Again, the textures derived from $A_i$ give a larger fraction of
predictions for the mixing angles and the ratio $r$ compatible with the data at the 3$\sigma$ level; the correlation among the matrix
elements is not as pronounced as in the other cases (especially for $\tilde C$, where the agreement to the experimental data is only possible 
at the prize of a huge conspiracy among the $\tilde a_i$ parameters).

In the inverted case, the situation is more puzzling; in fact, from the fit procedure we see that the pattern $C$ is less affected by fine-tuning and 
hierarchy problems than the $B_i$ textures (the $A_{1,2}$ patterns are in much worse agreement with the data and have not been further studied);
the approximate texture obtained from $C$ produces large mixing angles 
(with its matrix elements not suffering from strong correlations) and can be considered 
particularly suitable to describe large $\theta_{13}$.

\section*{Acknowledgements}  
We acknowledge MIUR (Italy) for financial support
under the program "Futuro in Ricerca 2010 (RBFR10O36O)".

\end{document}